\documentclass[
showpacs,
aps,
prb,
twocolumn,
superscriptaddress,
]{revtex4}

\usepackage{graphicx}
\usepackage{amsmath}
\usepackage{bm}
\usepackage{times}
\usepackage[varg]{txfonts}

\def\edc{\epsilon_{\mbox{\scriptsize dc}}}

\def\oc{\omega_{\mbox{\scriptsize {c}}}}

\def\rc{R_{\mbox{\scriptsize {c}}}}

\def\tq{\tau_{\mbox{\scriptsize {q}}}}
\def\ttr{\tau_{\mbox{\scriptsize {tr}}}}

\def\tqim{\tau_{\mbox{\scriptsize {q}}}^{\mbox{\scriptsize {im}}}}
\def\tqee{\tau_{\mbox{\scriptsize {q}}}^{\mbox{\scriptsize {ee}}}}

\begin{document}
\title{
Role of electron-electron interactions in nonlinear transport in 2D electron systems
}
\author{A.\,T. Hatke}
\affiliation{School of Physics and Astronomy, University of Minnesota, Minneapolis, Minnesota 55455, USA} 
\author{M.\,A. Zudov}
\affiliation{School of Physics and Astronomy, University of Minnesota, Minneapolis, Minnesota 55455, USA}
\author{L.\,N. Pfeiffer}
\affiliation{Bell Labs, Alcatel-Lucent, Murray Hill, New Jersey 07974, USA}
\author{K.\,W. West}
\affiliation{Bell Labs, Alcatel-Lucent, Murray Hill, New Jersey 07974, USA}
\received{February 21, 2009}
\accepted{April 07, 2009}

\begin{abstract}
We study the temperature evolution of the non-linear oscillatory magnetoresistance in a high-mobility two-dimensional electron system subject to a strong dc electric field.
We find that the decay of the oscillation amplitude with increasing temperature originates primarily from increasing quantum scattering rate entering the Dingle factor.
We attribute this behavior to electron-electron interaction effects.
\end{abstract} 
\maketitle

When a two-dimensional electron system (2DES) is subject to a weak perpendicular magnetic field $B$ and low temperature $T$, the linear response resistivity exhibits well-known Shubnikov-de Haas oscillations.\citep{shubnikov:1930}
These oscillations are controlled by the filling factor $\nu=2E_F/\hbar\oc$ ($E_F$ is the Fermi energy and $\hbar \oc$ is the cyclotron energy), and are periodic in $1/B$:
\begin{equation}
\Delta \rho_{\rm SdHO}=4\rho_0 \frac {X_T}{\sinh X_T} \delta \cos(\pi\nu)
\label{sdho}
\end{equation}
Here, $\rho_0$ is the resistivity at zero magnetic field, $X_T=2\pi^2  T/\hbar\oc$, and $\delta=\exp(-\pi/\oc\tq)$ is the Dingle factor.
From the dependence of the oscillation amplitude on magnetic field and temperature, one can deduce the quantum scattering time $ \tq$ and the effective mass $m^*$ of the charge carrier.

Recently, several other types of low-field magnetoresistance oscillations have been discovered in 2DES.
Among these are microwave-induced resistance oscillations,\citep{zudov:2001a,ye:2001} phonon-induced resistance oscillations,\citep{zudov:2001b} and Hall field-induced resistance oscillations,\citep{yang:2002a} which appear when 2DES is subject to microwaves, elevated (a few Kelvin) temperatures, or dc electric field, respectively (or a combination of microwave and dc electric fields\citep{zhang:2007c,hatke:2008,hatke:2008b}).
All these oscillations are also periodic in $1/B$ but the effective mass is available directly from the oscillation frequency.
Remarkably, microwave and dc fields can drive the 2DES into a state with zero resistance\citep{mani:2002,zudov:2003,yang:2003,zudov:2004,willett:2004,du:2004b,smet:2005,mani:2005,yang:2006,zudov:2006a,zudov:2006b,bykov:2006,dorozhkin:2009,andreev:2003,auerbach:2005,finkler:2006,finkler:2009} or zero differential resistance.\citep{bykov:2007,zhang:2008,romero:2008}

Stepping from inter-Landau level transitions all induced oscillations rely on both initial, $\nu(\varepsilon)$, and final, $\nu(\varepsilon+\Delta\varepsilon)$, densities of states. 
Here, $\Delta\varepsilon$ is the energy provided by microwave photon, acoustic phonon, or dc electric field.
In the regime of overlapped Landau levels, the density of states is given by $\nu(\varepsilon) = \nu_0[1-2\delta \cos(2\pi \varepsilon/\hbar\oc)]$, where $\nu_0$ is the density of states at zero magnetic field. 
In contrast with Shubnikov-de Haas oscillations, the leading (oscillating with $\Delta \varepsilon/\hbar\oc$) contribution to the resistivity originates from the $\delta^2$ term generated by the product of the oscillatory parts of the corresponding densities of states. 
This term survives averaging over the Fermi distribution, $\langle \cos^2(2\pi\varepsilon/\oc) \rangle_\varepsilon \simeq 1/2$, and therefore, unlike in the Shubnikov-de Haas effect, the temperature smearing of the Fermi surface does not come into play. 
As a result, all $\delta^2$ oscillations persist to a considerably higher temperature compared to Shubnikov-de Haas.
However, once the temperature is raised above a few Kelvin, oscillations start to decay rather rapidly.
It is therefore important to examine the temperature evolution of these oscillations and identify possible mechanisms responsible for their high temperature decay.

Recently, temperature dependence of microwave-induced resistance oscillations was examined by two experimental groups. 
In a first series of experiments,\citep{studenikin:2005,studenikin:2007} it was found that the oscillation amplitude decays as $T^{-2}$.
More recent studies\citep{hatke:2009a} using higher-mobility structures found the exponential decay of the amplitude, $\exp(-\alpha T^2)$ with $\alpha \propto 1/B$. 
This observation was explained in terms of electron-electron scattering which becomes relevant in high mobility 2DES at just a few Kelvin.
As far as phonon-induced resistance oscillations are concerned, their amplitude becomes vanishingly small at low temperatures due to lack of energetic acoustic phonons.
However, the decay at higher temperatures can also be linked to electron-electron interaction effects.\citep{hatke:2009b}

In this Communication we study the temperature evolution of Hall field-induced resistance oscillations in high-mobility 2DES which was not experimentally examined to date.
Our results show that the main source of the decay with increasing temperature is the decrease of the quantum scattering time entering the square of the Dingle factor.
We further find that the temperature-induced correction to the quantum scattering rate grows roughly as $T^2$ and thus can also be attributed to the electron-electron interaction effects.

Measurements were performed in a $^{3}$He cryostat on multiple lithographically defined Hall bars fabricated from symmetrically doped GaAs/Al$_{0.24}$Ga$_{0.76}$As quantum wells.
All the data presented here are from a $100$ $\mu$m-wide specimen with the density $n_e\simeq$ $3.7 \times 10^{11}$ cm$^{-2}$ and the mobility $\mu\simeq 1.0 \times 10^7$ cm$^2$/Vs, obtained after brief low-temperature illumination with red light-emitting diode.
The differential resistivity $r=dV/dI$ was recorded at temperatures from 2.0 to 5.0 K under applied constant current $I=80$ $ \mu$A in sweeping magnetic field using a low frequency (a few Hz) lock-in detection.

\begin{figure}[t]
\includegraphics{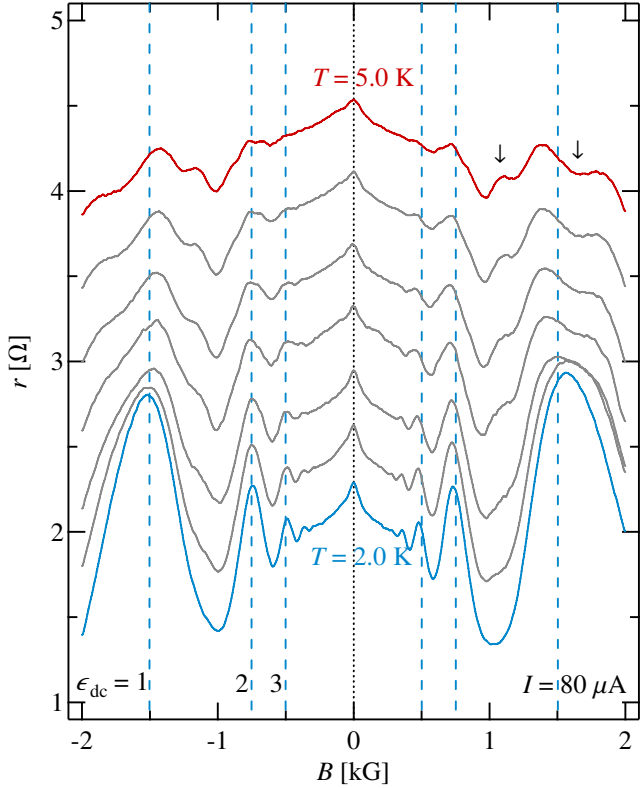}
\caption{(color online) Differential magnetoresistivity $r(B)$ at $I=80 \mu$A for temperature from 2.0 K to 5.0 K in 0.5 K steps.
The traces are vertically offset for clarity by 0.2 $\Omega$.
Vertical lines ($\edc=1,2,3$) mark oscillation maxima.}
\label{tdata}
\end{figure}
To explain nonlinear resistivity in high Landau levels of 2DES two physical mechanisms were theoretically considered.
One, commonly referred to as the ``displacement'' mechanism, is based on large-angle
scattering off of short-range disorder potential. 
Another mechanism, known as ``inelastic'',\citep{dorozhkin:2003,dmitriev:2005,dmitriev:2007,vavilov:2007} stems from the oscillatory dc-induced correction to the electron distribution function.
It was found\citep{vavilov:2007} that the ``inelastic'' mechanism is important only at very weak electric fields and thus cannot account for the oscillations.
On the other hand, the ``displacement'' mechanism\citep{lei:2007,vavilov:2007} provides excellent description of the experimental results.
Within this model oscillations in differential resistivity $r$ originate from elastic impurity-assisted electron transitions between Hall field-tilted Landau levels.
Oscillations are governed by a parameter $\edc=eE(2\rc)/\hbar\oc$, where $E$ is the Hall electric field and $2\rc$ is the cyclotron diameter, and at $\edc \gtrsim 1$, are described by:\citep{vavilov:2007,khodas:2008}
\begin{equation}
\frac {\Delta r} {\rho_0} =  \frac{(4\delta)^2}{\pi} \frac{\ttr}{\tau_{\pi}} \cos(2\pi\edc).
\label{theory}
\end{equation}
Here $\ttr$ is the impurity contribution to the transport scattering time and $\tau_{\pi}$ is the time describing electron backscattering off of impurities.
As discussed above, Hall field-induced resistance oscillations are insensitive to the temperature smearing of the Fermi surface and the temperature damping factor $X_T/\sinh X_T$ does not appear in Eq.\,(\ref{theory}). 
On the other hand, Hall field-induced resistance oscillations, appearing in the second order of the Dingle factor, should be more sensitive to disorder and thus call for 2DES with very long $\tq$.
However, contrary to what one might expect, Hall field-induced resistance oscillations typically persist down to much lower magnetic fields compared to Shubnikov-de Haas oscillations, even at low temperatures.
This is because single-particle lifetime $\tq$ appearing in Eq.\,(\ref{theory}) usually exceeds that entering Eq.\,(\ref{sdho}) by at least a few times.
The latter can be understood by noting that Hall field-induced resistance oscillations rely on the local density of states and thus are insensitive to macroscopic density fluctuations which severely affect Shubnikov-de Haas oscillations exhibiting underestimated $\tq$.\citep{coleridge:1989}
\begin{figure}[t]
\includegraphics{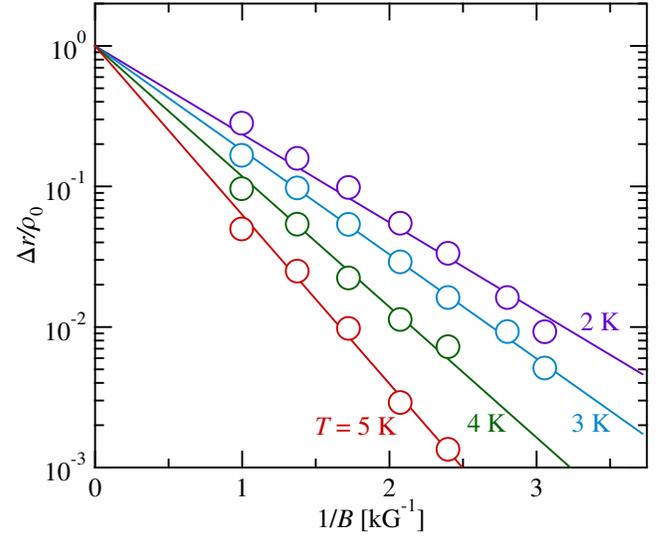}
\caption{(color online)
Circles show normalized oscillation amplitude $\Delta r/\rho_{0}$ vs $1/B$ for $T=$2, 3, 4, and 5 K. Lines represent fits to $\exp(-2\pi/\oc\tq)$.
}
\label{dingle}
\end{figure}

In Fig.\,\ref{tdata} we present the differential magnetoresistivity, $r(B)$ obtained at constant current $I=80$ $\mu$A for different temperatures from 2.0 K to 5.0 K, in 0.5 K increments.
The traces are vertically offset for clarity by 0.2 $\Omega$.
Vertical lines denote oscillation maxima found at $\edc\simeq 1,2,3$.\citep{zhang:2007a}
The lower temperature of our analysis is limited by Joule heating which affects the electron temperature at $\lesssim 2$ K.
We also note that Eq.\,(\ref{theory}) was derived assuming $T\gtrsim \hbar\oc$.
At higher temperatures, resonant acoustic phonon scattering distorts the oscillation waveform\citep{zhang:2008} (cf., $\downarrow$) limiting our ability to accurately extract the amplitude. 
Nevertheless, Fig.\,\ref{tdata} clearly shows that oscillations gradually decay with increasing temperature.
At the same time the zero-field differential resistivity shows monotonic growth which reflects the increase of linear resistivity $\rho_0$ due to enhanced scattering off of acoustic phonons.\citep{stormer:1990,mendez:1984}

The temperature dependence of the oscillation amplitude in Eq.\,(\ref{theory}) can originate from several parameters.
One is the Drude resistivity $\rho_{0}$ which exhibits monotonic growth with temperature and thus cannot be the cause of the decay.
We will use it as a normalizing factor in our analysis to correct for the change in the background resistance.
Another parameter is the ratio $\ttr/\tau_{\pi}$ which, however, should be treated as temperature independent since it only reflects the type of disorder.\citep{khodas:2008} 
Therefore the main candidate for the temperature dependence is the Dingle factor squared $\delta^{2}$ which contains $\tq$ and we continue our analysis by constructing Dingle plots.

From the data shown in Fig.\,\ref{tdata} we extract the normalized oscillation amplitudes $\Delta r/\rho_{0}$ and present the results in Fig.\,\ref{dingle} as a function of inverse magnetic field for $T=$ 2.0, 3.0, 4.0, and 5.0 K.
One immediately observes that the oscillation amplitude exhibits expected exponential decay over at least an order of magnitude.
In addition, one also observes that the exponent monotonically grows by absolute value with increasing temperature signaling a considerable decrease of the single-particle lifetime $\tq$.
At the same time, the extrapolation of all the data to $B^{-1}=0$ converge to a single point, in agreement with Eq.\,(\ref{theory}).
From the value of this intercept we obtain an estimate for the backscattering rate, $\tau_{\pi}^{-1} \simeq 0.18 \ttr^{-1}$. 

\begin{figure}[t]
\includegraphics{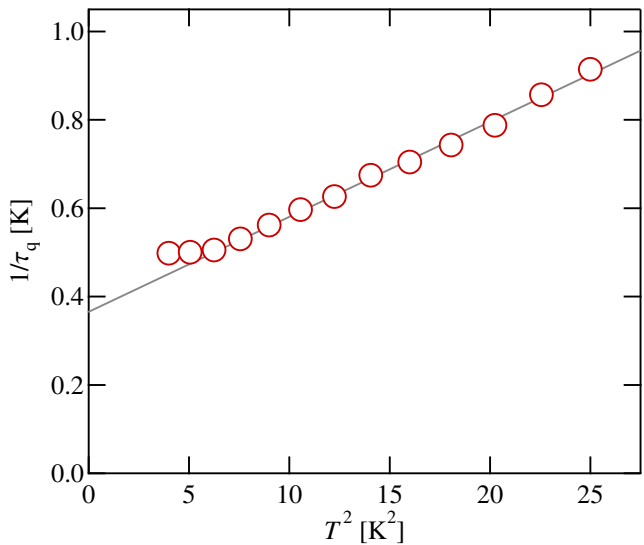}
\caption{(color online)
Quantum scattering rate $1/\tq$ vs $T^{2}$ (open circles) and linear fit (solid line).
}
\label{tauq}
\end{figure}
We repeat the Dingle plot analysis for all other temperatures studied and present the extracted quantum scattering rate $1/\tq$ in Fig.\,\ref{tauq} as a function of $T^2$.
One observes that the extracted quantum scattering rate increases roughly as $T^2$ over most of the temperature range. 
Such temperature dependence was recently obtained from the analysis of intersubband magnetoresistance oscillations in double quantum wells\citep{mamani:2008} and in studies of microwave-induced\citep{hatke:2009a} and phonon-induced\citep{hatke:2009b} resistance oscillations in single quantum wells.
In all cases, such characteristic dependence was viewed as a signature of electron-electron interaction effects. 
Employing Matthiessen's rule, we write $1/\tq=1/\tqim+1/\tqee$, where $1/\tqim$ is the temperature-independent impurity contribution and $1/\tqee$ is the electron-electron contribution.
Further assuming\citep{chaplik:1971,giuliani:1982} $1/\tqee=\lambda T^{2}/E_F$ where $\lambda \sim 1$ we fit our data and obtain $\lambda\simeq 4.1$ and $\tqim \simeq 20$ ps.
The obtained value of $\lambda$ is in good agreement with that obtained from the analysis of the temperature evolution of microwave-induced\citep{hatke:2009a} and phonon-induced\citep{hatke:2009b} resistance oscillations.
One can notice that at temperatures below $\simeq 2$ K the quantum scattering time tends to saturate at $\simeq 16$ ps.
We qualitatively explain the low-temperature departure from the $T^{2}$-dependence by Joule heating from the applied dc current which raises the temperature of our 2DES above the measured bath temperature. 

\begin{figure}[b]
\includegraphics{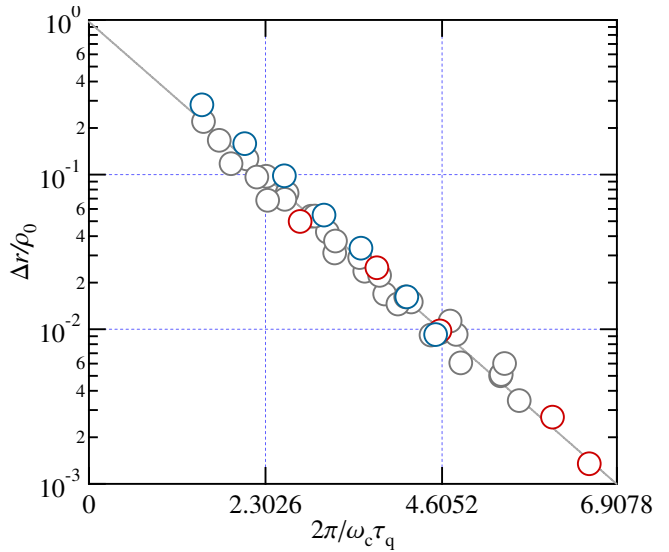}
\caption{(color online)
Circles show normalized oscillation amplitude $\Delta r/r_{0}$ vs $2\pi/\oc\tq$ for all temperatures studied.
Line marks a slope of $\exp(-2\pi/\oc\tq)$.
}
\label{alldata}
\end{figure}
To further confirm our observation we present in Fig.\,\ref{alldata} the normalized differential resistance as a function of $2\pi/\oc\tq$, where $1/\tq$ is computed using:
\begin{equation}
\frac{1}{\tq}=\frac{1}{\tqim}+
\lambda\frac{T^{2}}{E_F},
\label{tq}
\end{equation}
with the extracted values of $\tqim=20$ ps and $\lambda=4.1$
We observe that all our data obtained at different temperatures and magnetic fields closely follow a universal line prescribed by Eq.\,(\ref{tq}).
We thus conclude that the temperature dependence emerges primarily from quantum scattering rate modified by electron-electron interactions.

In summary, we have studied the temperature evolution of Hall field-induced resistance oscillations in a high mobility 2DES.
Our results show that the temperature dependence originates from the quantum scattering rate entering the square of the Dingle factor.
We find that this rate increases quadratically with increasing temperature which is a signature of electron-electron interactions.
Extracted electron-electron interaction scattering rate is in good agreement with recent experiments on microwave-induced\citep{hatke:2009a} and phonon-induced\citep{hatke:2009b} resistance oscillations in comparable mobility 2DES.
We thus conclude that the sensitivity to electron-electron scattering is a generic property of low-field magnetoresistance oscillations which appear in the second order of the Dingle factor.
This is in contrast to Shubnikov-de Haas oscillations, which, to the first order, are insensitive to electron-electron interactions.\citep{martin:2003,adamov:2006}

We thank I. A. Dmitriev, M. Dyakonov, M. Khodas, A. D. Mirlin, D. G. Polyakov, B. I. Shklovskii, and M. G. Vavilov for useful discussions. 
The work at the University of Minnesota was supported by NSF Grant No. DMR-0548014.


\end{document}